\documentclass{article}

\usepackage{arxiv}

\usepackage[utf8]{inputenc} 
\usepackage[T1]{fontenc}    
\usepackage{hyperref}       
\usepackage{url}            
\usepackage{booktabs}       
\usepackage{amsfonts}       
\usepackage{nicefrac}       
\usepackage{microtype}      

\usepackage{graphicx}
\usepackage{natbib}
\usepackage{doi}

\title{Seamless Integration: The Evolution, Design, and Future Impact of Wearable Technology}

\author{ \href{https://orcid.org/0009-0000-1518-3565}{\includegraphics[scale=0.06]{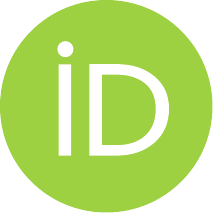}\hspace{1mm}David Pearl}\\
	Department of Mechanical Engineering\\
	Tufts University\\
	Medford, MA 02155 \\
	\texttt{David.Pearl@tufts.edu} \\
	\And
	\href{https://orcid.org/0000-0002-5487-2715}{\includegraphics[scale=0.06]{orcid.pdf}\hspace{1mm} Dr. James Intriligator} \thanks{Corresponding author} \\
	Department of Mechanical Engineering\\
	Tufts University\\
	Medford, MA 02155 \\
	\texttt{James.Intriligator@tufts.edu} \\
	\And
	\href{https://orcid.org/0000-0001-8305-2436}{\includegraphics[scale=0.06]{orcid.pdf}\hspace{1mm}Xuanjiang Liu} \\
	Department of Mechanical Engineering\\
	Tufts University\\
	Medford, MA 02155 \\
	\texttt{Xuanjiang.Liu@tufts.edu} \\
}

\renewcommand{\headeright}{Whitepaper}
\renewcommand{\undertitle}{Whitepaper}
\renewcommand{\shorttitle}{Seamless Integration}

\hypersetup{
pdftitle={Seamless Integration: The Evolution, Design, and Future Impact of Wearable Technology},
pdfsubject={q-cs.HC},
pdfauthor={David Pearl, James Intriligator, William Liu},
pdfkeywords={Wearable Technology, Human-Computer Interaction (HCI), Artificial Intelligence (AI), Internet of Things (IoT), Augmented Reality (AR), User-Centered Design, Data Privacy, Sustainability, Healthcare, Future Trends}}

\begin{document}
\maketitle

\begin{abstract}
The rapid evolution of wearable technology marks a transformative phase in human-computer interaction, seamlessly integrating digital functionality into daily life. This paper explores the historical trajectory, current advancements, and future potential of wearables, emphasizing their impact on healthcare, productivity, and personal well-being. Key developments include the integration of artificial intelligence (AI), Internet of Things (IoT), and augmented reality (AR), driving personalization, real-time adaptability, and enhanced user experiences. The study highlights user-centered design principles, ethical considerations, and interdisciplinary collaboration as critical factors in creating wearables that are intuitive, inclusive, and secure. Furthermore, the paper examines sustainability trends, such as modular designs and eco-friendly materials, aligning innovation with environmental responsibility. By addressing challenges like data privacy, algorithmic bias, and usability, wearable technology is poised to redefine the interaction between humans and technology, offering unprecedented opportunities for enrichment and empowerment in diverse contexts. This comprehensive analysis provides a roadmap for advancing wearables to meet emerging societal needs while fostering ethical and sustainable growth.

\end{abstract}


\section{Introduction}

\subsection{The Wearable Revolution: From Science Fiction to Everyday Life}

As we enter 2025, wearable technology is no longer futuristic—for many it is an integrated part of daily life, shaping how we interact with our environment from the moment we wake up. Imagine this: It is 2030, and your morning begins with technology seamlessly integrated into your routine. As you lace up your smart sneakers, their cushioning adjusts automatically based on your weight, gait, and the terrain ahead \citep{s_smart_2021}. Your smartwatch, fitted with a germ-resistant, customizable band, buzzes gently, reminding you to take your vitamins before you leave the house \citep{baek_reimagining_2024,himi_medai_2023}. As you jog, augmented reality (AR) glasses project your path onto the road and display real-time performance metrics while offering personalized coaching \citep{chang_a2fitnessartificial_2023}. This is not science fiction—it is the rapidly approaching future of wearable technology.
The evolution of wearable technology has been shaped by advances in miniaturization, wireless connectivity, and sensor technology, transforming early calculator watches of the 1980s into today’s sophisticated smart devices. Today, these devices integrate advanced sensors, artificial intelligence (AI), network connectivity, and augmented reality to enhance daily life \citep{ometov_survey_2021}. From fitness trackers monitoring heart rate and sleep patterns to smartwatches enabling seamless connectivity, wearables are becoming indispensable tools for health, safety, productivity, and convenience \citep{patel_trends_2022}. 
Yet their impact goes far beyond these functionalities. Wearables have the potential to revolutionize healthcare, enhance productivity, and enrich or even save lives. Smartwatches, for instance, are already capable of detecting arrhythmias and alerting emergency services in critical moments \citep{bogar_detection_2024}. In surgery, AR glasses are guiding procedures, reducing risks, improving precision, and minimizing complications \citep{dennler_augmented_2021}. Imagine the impact of a prenatal band that the parent wears to monitor detailed real-time states of the developing embryo—offering recommendations and insights \citep{alim_wearable_2023}. As technologies evolve, we will see wearables that augment everything from physical to cognitive and emotional aspects of our lives. The possibilities are vast, with wearables poised to transform how we interact with technology and the world around us.

\subsection{The Power of AI and Emerging Technologies}

Artificial intelligence (AI)—especially machine learning and LLM technology—is a driving force behind the next wave of wearables. By analyzing vast amounts of data, AI-powered devices can learn and adapt to user behavior. For example, fitness trackers now suggest indoor exercises when they detect inclement weather patterns in a user’s locale \citep{das_strength_2017}. Generative AI enhances personalization by enabling wearables to dynamically adapt interfaces and functionalities based on a user’s activity patterns, cognitive state, and aesthetic preferences, fostering a truly individualized experience \citep{shahhosseini_efficient_2022}.
Emerging technologies like the Internet of Things (IoT), 5G networks, and blockchain are enhancing connectivity and security. Advancements in AI and IoT enable smartwatches to function as interconnected nodes within broader digital ecosystems, facilitating real-time interaction with smart homes, vehicles, and healthcare networks to anticipate and respond to user needs \citep{dwivedi_blockchain_2024}. These advances promise to make wearables integral to our daily lives, seamlessly blending digital functionality with physical experiences.

\subsection{Designing for Humans: The Importance of User-Centered Design}

While technological advancements are pushing wearables headlong into an exciting future, human users remain a central focus of the design process. Wearables must be intuitive, comfortable, and tailored to the diverse and changing needs of individuals \citep{ferraro_designing_2011}. User-centered design ensures that these devices address real-world challenges while remaining functional and delightful. For example, smartwatches designed for seniors often feature large, high-contrast text and simplified interfaces to enhance usability \citep{while_glanceable_2024}. Similarly, wearables for children prioritize safety, durability, and playful interactions \citep{benisha_design_2021}. By considering ergonomics, cultural contexts, and emotional connections, designers can create devices that feel like extensions of the user rather than intrusive gadgets.
Beyond user experience, wearable design must incorporate broader considerations spanning political, economic, social, technological, environmental, and legal domains to ensure ethical and sustainable innovation. 

\subsection{Navigating the Future of Wearables}

This paper explores the evolution of wearable technology, key design methodologies, ethical and social implications, and future trends shaping the industry. By examining historical developments, user-centered design principles, and emerging innovations, we provide a comprehensive guide for engineers, designers, and industry leaders navigating this dynamic field.

\section{Foundations and Evolution of Wearable Technology}
\subsection{The Early Days: From Calculators to Fitness Trackers}

The beginnings of wearable technology can be traced to the 1970s with the development of digital wristwatches, which integrated fundamental computational capabilities alongside traditional timekeeping functions \citep{boettcher_invention_2023}. Although often clunky and not entirely user-friendly, these pioneering devices established foundational principles for integrating technology into wearable form factors. Advances in microprocessor technology throughout the 1980s and 1990s enabled wearables to adopt more sophisticated capabilities, setting the stage for broader adoption \citep{ometov_survey_2021}.
The 1990s and early 2000s saw the emergence of fitness-focused wearables, such as the Polar Heart Rate Monitor and Nike+iPod Sports Kit. These devices provided real-time insights into physical activity and heart rate, helping users optimize workouts and track progress \citep{patel_review_2012}. By demonstrating the potential for wearables to enhance personal health and wellness, while fitting seamlessly into the user’s lifestyle, these products laid the groundwork for the fitness wearables market that dominates today.

\subsection{The Smartphone Era and the Rise of Smartwatches}

The proliferation of smartphones in the late 2000s—along with standardized and low-energy forms of wireless communication (e.g., Bluetooth)—catalyzed a major shift in wearable technology. As consumers became accustomed to carrying powerful computing devices, wearables evolved to complement and extend smartphone functionality \citep{seneviratne_survey_2017}. This convergence of technologies reshaped user expectations, paving the way for a new generation of interconnected devices.
Smartwatches epitomize this shift. With the launch of the Pebble smartwatch (2013) and the Apple Watch (2015), wearables evolved into multifunctional hubs that seamlessly integrated smartphone capabilities, including notifications, applications, and biometric tracking. These wearables introduced novel interaction paradigms, such as glanceable displays, voice commands, and haptic feedback, redefining how users interacted with technology on the go \citep{mcmillan_situating_2017}. By merging utility and convenience, smartwatches cemented their place as versatile tools in the digital ecosystem.

\subsection{The Expansion of Wearable Applications}

As the technology matured, wearables moved beyond fitness and wellness into diverse applications. In healthcare, wearables such as continuous glucose monitors enable remote patient monitoring, providing real-time data to physicians and empowering patients to manage chronic conditions \citep{alim_wearable_2023}. Assistive technologies, such as wearable navigation aids, have expanded accessibility for individuals with disabilities \citep{pellecchia_how_2019}. 
In professional settings, wearable technologies have transformed industries like logistics and manufacturing. Augmented reality (AR) headsets, such as Microsoft’s HoloLens, provide hands-free access to complex procedures and workflows, enhancing productivity and safety \citep{plakas_augmented_2020}. Smart gloves, wearable scanners, and wearable exoskeletons streamline operations in warehouses, illustrating the potential for wearables to improve efficiency, safety, and precision across sectors \citep{kinoshita_wearable_2018}.

\subsection{The Future of Wearables: Integration and Personalization}

The future of wearables lies in deeper integration with emerging technologies (sensors, actuators, and AI) and a focus on personalized experiences. As components of the Internet of Things (IoT), wearables will increasingly interact with smart homes, vehicles, urban environments, and other wearables (personal or those of others), enabling responsive, context-aware applications \citep{john_dian_wearables_2020}. Advances in AI and machine learning will further enhance this connectivity, allowing devices to adapt to user preferences and offer proactive recommendations \citep{seng_machine_2023}.
Personalization is also transforming wearable design. Innovations in materials science, such as flexible electronics and smart textiles, are enabling wearables that conform to individual anatomy and aesthetics \citep{azani_electronic_2024}. Technologies like 3D printing facilitate custom-fit designs, blurring the line between device and user \citep{rajendran_revolutionizing_2024}. These advancements suggest a future where wearables are not merely tools but seamless extensions of personal identity and lifestyle.

From the rudimentary digital watches of the 1970s to the sophisticated, AI-powered devices of today, wearable technology has evolved in response to changing user needs and technological landscapes. As these devices continue to integrate with emerging technologies and prioritize personalization, they are poised to redefine how we live, work, and interact. Yet, this progress also demands thoughtful consideration of ethical and societal implications. By addressing challenges such as privacy and accessibility, the next generation of wearables can maximize their potential to enrich human experiences responsibly.

\section{Design Philosophies and Collaborative Approaches}
\subsection{Putting the "Human" in Human-Centered Design}

In wearable technology design, the most critical component is not found in a sensor or processor—it is the human who wears it. Human-centered design (HCD) and user-centered design (UCD) place users at the heart of the design process, ensuring wearables are not only functional but also intuitive, accessible, enriching, and enjoyable to use \citep{noauthor_iso_2019}.
Consider the Apple Watch. Apple's design process involved extensive user observation to understand how people interacted with traditional watches. This understanding was then combined with detailed analysis of functions that could benefit from real-time interactions when a watch was the most convenient device. For example, when navigating an urban center, simple haptic directions delivered via the watch provide a less intrusive and more convenient way of delivering directions \citep{apple_inc_designing_nodate}. The result was a device that seamlessly integrates fitness tracking, health monitoring, navigation, and communication into a premium, user-friendly timepiece. This approach exemplifies how a deep understanding of user needs can transform technology into a valuable daily companion.

\subsection{Breaking Down Silos: The Power of Interdisciplinary Collaboration}

Wearable design requires input from multiple disciplines: engineers developing the hardware, designers crafting functions and user interfaces, data scientists analyzing sensor inputs and outputs, and healthcare experts ensuring safety and efficacy. Collaboration across these fields is essential to creating devices that are both functional and user-friendly \citep{nguyen_dimensions_2020}.
The Empatica Embrace, a wearable that detects seizures, highlights the success of this approach. By bringing together experts in neurology, data science, engineering, and design, the team developed a device that accurately detects seizures while maintaining the aesthetic appeal of a stylish smartwatch \citep{noauthor_embrace2_nodate}. This synergy of expertise ensured both technical precision and user acceptance.

\subsection{Designing for All: Inclusive and Ethical Design}

Wearable technology must cater to diverse users, including those with varying abilities, ages, and cultural contexts. Inclusive design prioritizes accessibility and usability, ensuring that wearables accommodate diverse demographics, including individuals with varying cognitive, sensory, and physical needs. Inclusive design ensures accessibility through features like adjustable fonts, haptic feedback, and modular components \citep{hokka_gender_2023}. Ethical design goes further, addressing data privacy, fairness, and transparency in algorithmic decision-making \citep{tsamados_ethics_2022}.
A cautionary example is the Strava fitness app, which inadvertently revealed sensitive military locations through its global heatmap feature. This incident underscores the importance of ethical considerations, such as limiting data sharing and conducting rigorous impact assessments \citep{reed_fitness_2018}. Designers must anticipate unintended consequences and prioritize user well-being over convenience or profit.

\subsection{Balancing Innovation and Practicality}

Wearable design is a balancing act between pushing technological boundaries and creating devices people will use daily. Innovation must be tempered with practicality, ensuring devices are comfortable, affordable, and relevant to real-world contexts.
The Oura Ring exemplifies this balance. By focusing on sleep and recovery metrics rather than incorporating every possible feature, the Oura team created a sleek, lightweight device with long battery life and actionable insights. This focus on user experience has earned the Oura Ring a devoted following among athletes and health enthusiasts \citep{noauthor_oura_nodate}.
Designing wearable technology requires more than technical expertise—it demands empathy, collaboration, and a commitment to ethical principles. By prioritizing human needs, embracing interdisciplinary teamwork, and balancing innovation with practicality, we can create wearables that truly enhance lives. In the next section, we will explore the core methods and frameworks that turn these philosophies into actionable designs.

\section{Core Design Methods and Approaches}
\subsection{User Research and Insight Gathering}

At the heart of any successful wearable technology lies a deep understanding of the user. As the renowned designer Charles Eames once said, "The role of the designer is that of a very good, thoughtful host anticipating the needs of his guests" \citep{noauthor_charles_nodate}. In the realm of wearables, this means employing a range of user research methods to uncover insights that will guide the design process. In this section we provide a brief discussion and roadmap of some of the most common and central research methods that can help inform the design of successful wearables.

\subsubsection{Contextual Inquiry and Ethnographic Research}

Contextual inquiry and ethnographic research are foundational methods in human-centered design, allowing researchers to study user behaviors, needs, and challenges within real-world environments. Unlike traditional user testing, which often occurs in controlled settings, these methods focus on capturing authentic interactions by embedding researchers within the user's daily experiences.
Contextual inquiry is a semi-structured field interview method where researchers observe users as they engage with a product or process while simultaneously asking clarifying questions. The goal is to uncover implicit behaviors, pain points, and workarounds that may not surface in lab-based studies. For example, when investigating the use of wearable glucose monitors by individuals with diabetes, researchers conducting contextual inquiries may shadow users in their homes or workplaces to understand how they check their glucose levels, what obstacles they encounter, and how they integrate the device into their routines.
Ethnographic research takes a broader, more immersive approach. It involves prolonged engagement with users in their natural contexts—whether at home, work, or in public spaces—to gain deeper insights into cultural, social, and behavioral influences on technology use. This method is particularly valuable for understanding long-term adoption patterns and social perceptions of wearables. In a study on fitness tracker use among older adults \citep{kononova_use_2019}, ethnographic research revealed that many seniors struggled with small text and complex navigation. These findings led to design recommendations such as larger displays, simplified interfaces, and voice-guided instructions to improve usability and adoption.
By leveraging these qualitative research methods, designers can develop wearables that align more seamlessly with real-world needs, ensuring usability, accessibility, and long-term engagement across diverse user groups.

\subsubsection{Observational Methods (Structured and Unstructured)}

Observational methods, both structured and unstructured, provide valuable insights into how users interact with wearables in various settings. Structured observations involve predefined criteria and focus on specific behaviors, while unstructured observations allow for a more open-ended exploration of user actions and reactions. 
In a series of studies investigating how users interact with smartwatches in public spaces, the authors used structured observations to gain deeper insights \citep{mcmillan_situating_2017, pizza_smartwatch_2016}. The researchers noted that users often struggled to navigate the small screens and frequently had to reorient the device to read notifications, leading to awkward and conspicuous interactions. These observations suggested a need for more discreet and intuitive interfaces that minimize the need for overt physical manipulations.

\subsubsection{Surveys, Questionnaires, Interviews, and Focus Groups}

Traditional research methods like surveys, questionnaires, interviews, and focus groups remain essential tools for gathering both quantitative and qualitative data on user preferences, behaviors, and experiences with wearables.
For instance, \citet{maher_users_2017} conducted a cross-sectional study involving 237 Australian adults who were current or former users of wearable activity trackers. Their findings revealed that users highly valued features such as step counting and heart rate monitoring, while technical issues like battery life and data privacy concerns were commonly reported. These insights highlight the importance of addressing both functional and emotional aspects of the user experience when designing wearables.

\subsubsection{Cultural Probes}

Cultural probes are a creative, exploratory method for gathering rich, qualitative data about users' cultural and emotional contexts. By providing users with open-ended tasks or prompts, designers can elicit reflections and insights that might not emerge through more structured research methods.
In a study exploring the design of emotionally engaging wearables, \citet{silina_new_2015} used cultural probes to understand how users might relate to wearable devices on a personal level. Participants were given a kit containing a diary, a disposable camera, and various prompts encouraging them to reflect on their daily experiences and emotions. The resulting data revealed a strong desire for wearables that could adapt to users' changing moods and needs, leading to design concepts for devices that could sense and respond to emotional states.

\subsection{Design and Prototyping}
\subsubsection{Design Thinking (EDIPT Framework)}

Design thinking is a human-centered approach to innovation that has been widely adopted in wearable technology development. The EDIPT framework—Empathize, Define, Ideate, Prototype, Test—offers a structured process for understanding user needs, generating solutions, and iterating based on feedback. According to \citet{venkatesh_investigating_2022}, this framework facilitates a deeper understanding of user expectations and enables designers to develop solutions that effectively bridge the gap between these expectations and technological possibilities.
One notable application of design thinking is the development of the Embrace smartwatch for seizure detection by Empatica. Their design process started with empathizing with epilepsy patients and caregivers to understand the challenges they face in managing the condition. They then defined the need for a discreet, comfortable device to detect seizures and alert loved ones. Through ideation sessions, the team prototyped and tested solutions with users, culminating in the Embrace watch, which has been praised for its user-centered design and ability to improve the lives of individuals with epilepsy \citep{noauthor_embrace2_nodate}.

\subsubsection{Advanced Design Models: Transversal Design}

While traditional frameworks like EDIPT provide a structured foundation for user-centered design, Transversal Design offers a more expansive, multidimensional approach that better addresses the complexity of wearable technology. This methodology encourages designers to move beyond conventional problem-solving and instead “glide” across diverse dimensions, uncovering constraints, opportunities, and synergies that would otherwise go unnoticed.
Unlike traditional design thinking, which primarily focuses on user interaction, emotional experience, and function, Transversal Design opens the design space to a near-infinite range of considerations. It challenges designers to cut across disciplines, integrating insights from fashion, sustainability, behavioral psychology, biomechanics, real-time data ethics, and emerging technologies—all within a single design exploration. For example, a health-tracking wearable would not just prioritize usability and comfort but would also intersect with sustainability (eco-friendly materials), social perception (how wearables influence self-image), and AI-driven personalization (adaptive biometric feedback).
A key feature of Transversal Design is its emphasis on interdisciplinary collaboration. Instead of a sequential design process, where technical feasibility follows aesthetic or functional considerations, this approach brings together fashion designers, engineers, behavioral scientists, ethicists, and data analysts from the outset. This ensures that the final product not only aligns with individual user needs but also responds to broader cultural, technological, and ethical shifts.
By leveraging artificial intelligence and computational design tools, Transversal Design enables more adaptive, future-proof wearable solutions. AI can assist designers in generating dynamic product variations, optimizing materials for sustainability, or simulating user interactions across diverse environments before a prototype is ever built. This integration of AI-powered design thinking with a transversal mindset pushes the boundaries of what wearables can achieve—ensuring that they evolve as more than just functional tools, but as dynamic, responsive extensions of human experience.

\subsubsection{Rapid Prototyping and Scenario-Based Design}

Rapid prototyping is integral to user-centered and transversal design, allowing iterative refinement based on empirical user feedback, thereby enhancing functional and experiential design aspects. Given the complexity of wearable technology—where factors such as ergonomics, interaction design, and sensor integration must seamlessly align—prototyping plays a crucial role in identifying usability challenges, optimizing functionality, and refining the user experience.
Prototyping in wearables can take many forms, ranging from low-fidelity paper sketches and foam models to mid-fidelity interactive wireframes and high-fidelity digital simulations. These iterations allow designers to explore everything from button placement and screen navigation to gesture-based interactions and sensor calibration, ensuring that wearables feel intuitive and effective in real-world use.

\subsubsection{Scenario-Based Design: Prototyping in Context}

One particularly effective strategy for wearable design is scenario-based prototyping, where designers create detailed, context-driven narratives to explore how users will interact with a device throughout their daily lives. Unlike static usability testing, scenario-based design immerses prototypes into dynamic, real-world environments, revealing hidden usability challenges and unforeseen design opportunities.
For example, when developing a fitness tracker, scenario-based prototyping might explore:
\begin{itemize}
    \item Morning: A user tracking their heart rate during a pre-work run in varying weather conditions.
    \item Workday: Using the device’s haptic feedback to remind them to take standing breaks.
    \item Evening: Transitioning the wearable to sleep tracking mode, assessing comfort and battery efficiency overnight.
    \item By framing wearable interactions within realistic user journeys, designers gain insights into device adaptability, comfort, accessibility, and seamless transitions between different modes of use.
\end{itemize}

\subsubsection{Innovative Prototyping Methods in Wearables}

Emerging fabrication methods are pushing the boundaries of what can be prototyped and how quickly iterations can be tested. \citet{hanton_protospray_2020} introduced ProtoSpray, an innovative method combining 3D printing and spray coating to create interactive displays on arbitrary surfaces. This approach enables rapid fabrication of custom-shaped wearable interfaces, moving beyond rigid screens to more flexible, body-contoured interaction surfaces. Through iterative prototyping and user testing, their team refined form factors, interaction methods, and material choices to enhance user experience.
Similarly, \citet{omaia_interactive_2024} explored hybrid prototyping, merging physical mockups with AI-driven digital simulations to predict real-world usability outcomes before manufacturing. This technique allows designers to assess gesture interactions, haptic feedback, and real-time biometric responses in a virtual prototyping environment, reducing time-to-market while ensuring optimal design performance.
By combining traditional prototyping with AI-enhanced simulations, rapid fabrication, and scenario-based evaluation, designers can create wearables that are not just functionally effective, but also deeply intuitive, contextually aware, and seamlessly integrated into users’ lives.

\subsection{Analysis and Refinement}
\subsubsection{Task Analysis in Wearable Design}

As wearable concepts take shape, designers must continually analyze and refine their designs to ensure they meet user needs and expectations. This involves a range of methods, from traditional usability testing and task analysis to more advanced techniques like multidimensional task analysis.

\subsubsection{Task Analysis and Multidimensional Task Analysis (MTA)}

Task analysis serves as a foundational method in wearable design, systematically deconstructing user interactions to identify inefficiencies, cognitive load, emotional reactions, and potential usability barriers. By breaking down complex tasks into their component steps, designers can identify potential bottlenecks, errors, and opportunities for improvement. In the context of wearables, task analysis often begins with a focus on physical interactions, such as how users navigate menus or activate features. However, as the devices become more sophisticated, there is a growing need to consider cognitive and emotional dimensions as well. 
This is where multidimensional task analysis (MTA) comes into play. MTA, as described by \citet{intriligator_multidimensional_2022}, is an approach that considers not only the physical actions users take but also their cognitive processes, decision points, social interactions, and affective states. By analyzing tasks across multiple dimensions, designers can create wearables that are more intuitive, engaging, and emotionally resonant. For example, in designing a wearable for stress management, an MTA approach might consider not only how users physically interact with the device (e.g., putting it on, adjusting settings) but also how they cognitively process the feedback it provides (e.g., interpreting stress levels, deciding on coping strategies) and how they interact with others (e.g. friends, family, therapists) and emotionally respond to the experience (e.g., feeling reassured, motivated, or frustrated). By addressing these multiple dimensions, designers can create a more holistic and effective solution.

\subsubsection{Anthropometric and Biomechanical Analysis}

Understanding human body measurements (anthropometry) and movement mechanics (biomechanics) is crucial in wearable design to ensure comfort, functionality, and user acceptance. For instance, designing a smart compression garment requires precise knowledge of body dimensions to achieve optimal fit and sensor placement. \citet{paiva_design_2018} discuss the design of smart garments for sports and rehabilitation, emphasizing the importance of integrating textile sensors into compression garments to monitor physiological parameters effectively. 
Additionally, compression garments have been shown to reduce muscle oscillation during exercise, contributing to greater stability and potentially enhancing performance. \citet{borras_effects_2011} demonstrated that compression gear could significantly decrease muscle oscillation during intense exercise, suggesting benefits for muscle stability and recovery with minimal associated risks.
By applying anthropometric and biomechanical analyses, designers can create wearables that not only fit well but also enhance user performance and comfort, leading to higher adoption rates and user satisfaction.

\subsubsection{Data-Driven Design and AI-Enhanced Insights}

As wearables become more sensor-rich and connected, they generate vast amounts of data on user behaviors, preferences, and biometrics. This data can be a goldmine for designers, providing unprecedented insights into how users interact with the devices and how they can be improved over time.
Data-driven design involves the systematic collection, analysis, and application of user data to inform design decisions. This can range from simple A/B testing of interface elements to more complex machine learning models that predict user needs and adapt the device's behavior accordingly.
For example, Oura Health, a company that produces a smart ring for sleep and activity tracking, uses data from its users to continuously refine its algorithms and provide more accurate and personalized insights. By analyzing patterns in sleep quality, heart rate variability, and other biometrics, the company can identify trends and correlations that inform updates to the device's firmware and companion app \citep{noauthor_oura_nodate}.
Similarly, researchers are developing AI-driven wearables that monitor emotional well-being. \citet{asif_proactive_2024} describe a proactive emotion tracker that integrates physiological signals from wearables, such as smartwatches and EEG sensors, with AI algorithms to provide continuous monitoring of emotional states. These technologies enable personalized interventions and recommendations, transforming how users manage their mental health and overall well-being.
AI and machine learning enhance the capabilities of wearables by training models on large datasets of user interactions and outcomes. This enables systems to adapt to individual users' needs and preferences over time, fostering a more personalized and engaging experience. Incorporating AI and data-driven methodologies in the design process facilitates the creation of intelligent wearables that not only offer actionable health insights but also adapt dynamically to user behaviors, ultimately increasing user satisfaction and improving health outcomes.

\subsection{Physical Design and Manufacturing}

While much of the focus in wearables design is on the digital and experiential aspects, the physical design and manufacturing of the devices are equally critical. The form factor, materials, and production processes used can have a significant impact on the device's comfort, durability, and overall user experience.

\subsubsection{Physical Forms and Ergonomics}

The physical form of a wearable device must strike a balance between functionality and comfort. Ergonomic considerations are paramount, as the device must fit seamlessly with the user's body and movements without causing discomfort or irritation.
One key consideration is the device's size and weight. Wearables that are too bulky or heavy can be cumbersome and distracting, while those that are too small may be difficult to interact with or may not have sufficient battery life. Designers must carefully balance these factors, taking into account the specific use case and target audience for the device \citep{verwulgen_new_2018}.
Another important aspect of ergonomic design is the device's shape and contours. Wearables that are designed to fit the natural curves of the body, such as the wrist or the ear, are generally more comfortable and secure than those with flat or rigid surfaces. This is particularly important for devices that are worn for extended periods, such as fitness trackers or smartwatches \citep{el-gayar_factors_2023}.
A case study by \citet{lacko_ergonomic_2017} illustrates the importance of ergonomic design in the development of a wearable EEG headset. Through a series of user studies and iterative prototypes, the researchers identified key design features that enhanced comfort and usability, such as a flexible, adjustable headband and strategically placed electrodes that minimized pressure points. The resulting device was not only more comfortable to wear but also provided more reliable and consistent EEG readings \citep{lacko_ergonomic_2017}).

\subsubsection{Materials and Durability}

The materials used in wearable devices must balance a range of factors, including durability, comfort, and aesthetics. As these devices are worn on the body and often exposed to the elements, they must be able to withstand regular wear and tear, as well as exposure to sweat, moisture, and other environmental factors.
Common materials used in wearables include plastics, metals, and elastomers. Each has its own strengths and weaknesses, and the choice of material will depend on the specific requirements of the device. For example, a fitness tracker designed for swimming will require a different set of materials than one designed for everyday wear \citep{liu_recent_2022}.
In addition to traditional materials, there is a growing interest in the use of smart materials and e-textiles in wearables design. These materials can sense and respond to various stimuli, such as temperature, pressure, or electrical signals, opening up new possibilities for interactive and adaptive wearables.
A study by \citet{gupta_soft_2021} explores the use of smart textiles in the design of a wearable knee brace for rehabilitation. The brace incorporates a conductive fabric sensor that can detect the wearer's knee joint angle and provide real-time feedback to guide their exercises. The use of a flexible, breathable fabric ensures that the brace is comfortable to wear for extended periods, while the integrated sensing capabilities enable more effective and personalized rehabilitation.

\subsubsection{Manufacturing Processes and Customization}

The manufacturing processes used to produce wearables can have a significant impact on their cost, quality, and customization options. Traditional manufacturing methods, such as injection molding and die casting, are well-suited for mass production but offer limited opportunities for customization. In recent years, there has been a growing interest in the use of additive manufacturing (AM) techniques, such as 3D printing, for wearables production. AM enables greater design freedom and allows for the creation of complex geometries and personalized features that would be difficult or impossible to achieve with traditional manufacturing methods \citep{wong_review_2012}.
One example of the use of AM in wearables is the Oura Ring, a smart ring that tracks sleep and activity metrics. The ring is 3D printed using a biocompatible polymer, allowing for a high degree of customization to fit each user's unique finger size and shape. This customization not only enhances comfort but also ensures a secure fit, which is critical for accurate sensor readings \citep{noauthor_oura_nodate}.
Another advantage of AM is the ability to rapidly prototype and iterate on designs. This is particularly valuable in the fast-paced world of wearables, where consumer preferences and technological capabilities are constantly evolving. By using AM to quickly test and refine designs, companies can bring new products to market faster and with greater confidence in their success.

\subsection{Interaction Design and Sensory Dimensions}
\subsubsection{Interaction Design (IxD) Principles}

Interaction Design (IxD) plays a crucial role in creating wearable interfaces that are intuitive, efficient, and user-friendly. Key IxD principles such as feedback, visibility, and consistency are essential for designing interfaces that make the most of limited screen real estate while providing a seamless user experience \citep{noauthor_what_2016}. 
Feedback mechanisms in wearable interfaces provide users with immediate, context-aware responses, enhancing interaction efficiency and user confidence. In the context of wearables, feedback can take many forms, from visual cues on the device's screen to haptic vibrations and auditory alerts. For example, a smartwatch might provide a subtle vibration to confirm that a message has been sent or display a green checkmark icon or happy melody to indicate a successful payment \citep{while_glanceable_2024}.
Visibility is another critical principle, particularly for wearables with small screens. It involves making essential features and information easily accessible and understandable. A well-designed smartwatch interface, for instance, might use clear, bold icons and simple navigation gestures to help users quickly access key functions without getting lost in complex menus \citep{islam_visualizing_2024}.
Consistency is also vital for creating intuitive wearable interfaces. It ensures that users can easily learn and navigate the device's interface by providing familiar and predictable patterns of interaction. For example, using similar gestures and button placements across different apps and functions can help users feel more confident and efficient when using their smartwatch \citep{apple_inc_designing_nodate}.
By adhering to these IxD principles, designers can create wearable interfaces that are not only visually appealing but also easily learnable, highly functional, and user-friendly. The Apple Watch is a prime example of effective IxD in action. Its interface uses simple, clear icons and intuitive navigation gestures, such as swiping and tapping, to help users efficiently access a wide range of features, from fitness tracking to mobile payments \citep{frances-morcillo_role_2018}.

\subsubsection{Glanceability and Screen Real Estate}

Glanceability is a critical consideration in the design of wearable interfaces, particularly for devices with small screens like smartwatches. It refers to the ability of users to quickly and easily access key information without having to perform complex interactions or navigate through multiple screens \citep{ferraro_designing_2011}.
Designing for glanceability involves prioritizing essential information and presenting it in a clear, concise manner. For example, a fitness tracker might display the user's current step count, heart rate, and active minutes on the main screen, allowing them to quickly check their progress throughout the day. More detailed information, such as historical data or settings, can be accessed through secondary screens or menus \citep{ferraro_designing_2011}.
Effective use of screen real estate is also crucial in wearable design. Given the limited space available, designers must carefully consider what information to display and how to lay it out for optimal readability and interaction. This often involves using high contrast, legible typography, and minimalistic UI elements to ensure the interface is visually clear and easy to navigate at a glance \cite{frances-morcillo_role_2018}. And, of course, always keeping the needs and desires of the specific end-users in mind (e.g. older adults vs kids).
Health alerts are a prime example of where glanceability and effective use of screen real estate are paramount. A well-designed smartwatch interface might use a combination of bold colors, simple icons, and concise text to communicate critical health information, such as an abnormal heart rate or a reminder to take medication. By presenting this information in a clear and instantly recognizable format, users can quickly understand the alert and take appropriate action without needing to navigate through complex menus or interpret ambiguous data \citep{frances-morcillo_role_2018}.

\subsubsection{Non-Visual Sensory Interactions (Haptics, Auditory)}

Beyond visual interfaces, wearables leverage multimodal interactions — haptic feedback, auditory signals, and adaptive sensory cues — to enhance usability and accessibility. These additional sensory channels often provide a richer, more immersive, and more compelling means of conveying information and alerts, particularly in situations where visual attention may be limited.
Haptic feedback, typically in the form of vibrations, can be used to communicate a wide range of information, from incoming notifications to navigation cues. By using distinct vibration patterns, wearables can convey different types of alerts without requiring users to look at the device. For example, a smartwatch might use a short, sharp vibration to indicate an incoming text message, while a longer, more sustained vibration could signal an incoming call \citep{hong_vibration-based_2023}.
Auditory cues, such as tones, beeps, and spoken prompts, can also provide valuable information and feedback to users. In the context of AR glasses, for instance, auditory cues can be used to provide turn-by-turn navigation instructions, freeing users' visual attention to focus on their surroundings. Similarly, a fitness tracker might use a celebratory jingle to indicate when a user has reached their daily step goal, providing a sense of achievement and motivation \citep{voigt-antons_impact_2024}.
The combination of haptic and auditory feedback can create a more engaging and intuitive user experience, allowing wearables to communicate with users in a more natural and immersive way. The Apple Watch, for example, uses a feature called the Taptic Engine, which provides a range of distinct haptic feedback patterns for different types of notifications and interactions. When combined with auditory cues, such as the subtle clicking sound of the Digital Crown, this creates a rich, multi-sensory experience that feels intuitive and responsive \citep{apple_inc_designing_nodate}.

\subsection{Usability and Evaluation}
\subsubsection{Usability Testing and Iterative Prototyping}

Usability testing is a critical step in the design process for wearable devices, allowing designers to identify potential issues and improve the overall user experience. By involving real users in the testing process, designers can gather valuable insights into how well the device meets users' needs and expectations, and identify areas for improvement \citep{andreoni_investigating_2023}.
Iterative prototyping is a key component of usability testing, involving the creation of multiple versions of the device or interface that can be tested and refined based on user feedback. This approach allows designers to quickly identify and address usability issues, ensuring that the final product is as user-friendly and effective as possible \citep{yoon_case_2017}.
When conducting usability tests for wearable fitness trackers, for example, it is important to involve a diverse group of users, including those with varying levels of technical expertise and fitness experience. By observing how these users interact with the device, designers can identify potential barriers to adoption, such as confusing navigation or unclear data displays. They can then use this feedback to refine the design, creating a more intuitive and accessible experience for all users \citep{singh_identifying_2019}.
Usability testing can also help designers identify potential safety and comfort issues, particularly for wearables that are designed to be worn for extended periods. For instance, testing a smartwatch with a diverse group of users can reveal issues with the device's fit or materials, such as straps that are too tight or cause skin irritation. By addressing these issues early in the design process, designers can create wearables that are not only functional but also safe and comfortable to wear \citep{seo_exploring_2024}.

\subsubsection{System Usability Scales (SUS) and Heuristic Evaluation}

In addition to user testing, there are several standardized tools and methods for evaluating the usability of wearable devices. The System Usability Scale (SUS) is a widely used questionnaire that provides a quick and reliable way to measure users' subjective perceptions of a device's usability \citep{alshamari_usability_2024, brooke_sus_1996}. SUS consists of ten statements that users rate on a five-point scale, ranging from strongly disagree to strongly agree. These statements cover a range of usability factors, such as ease of use, learnability, and user satisfaction. By calculating a composite score based on users' responses, designers can quickly gauge the overall usability of the device and compare it to industry benchmarks \citep{alshamari_usability_2024, brooke_sus_1996}.

Heuristic evaluation is another valuable tool for assessing the usability of wearable interfaces. This method involves having a group of usability experts review the device or interface against a set of established usability principles, known as heuristics. These heuristics cover a range of design factors, such as consistency, error prevention, and user control \citep{alshamari_usability_2024, nielsen_heuristic_1990}.
During a heuristic evaluation, experts identify potential usability issues and assign them a severity rating based on their impact on the user experience. For example, a smartwatch interface that uses inconsistent gestures or labels across different screens might be flagged as a high-severity issue, as it could significantly impede users' ability to navigate and use the device effectively \citep{alshamari_usability_2024}. \citet{guizerian_proposal_2024} have proposed a set of heuristics specifically for wearable devices.
By combining the results of SUS and heuristic evaluations with insights from user testing, designers can gain a comprehensive understanding of a wearable device's usability strengths and weaknesses. They can then use this information to prioritize design improvements and ensure that the final product meets the highest standards of usability and user-friendliness \citep{alshamari_usability_2024}.
As the wearable technology landscape continues to evolve, the importance of rigorous usability testing and evaluation cannot be overstated. By placing users at the center of the design process and continuously refining and improving the user experience, designers can create wearable devices that are not only technologically advanced but also truly intuitive and enjoyable to use.

\section{Ethical, Social, and Legal Considerations}

As wearable technology becomes increasingly integrated into our daily lives, it raises a host of ethical, social, and legal issues that must be carefully navigated. These devices, by their very nature, collect and process vast amounts of personal data, from our physical movements to our physiological metrics and even our emotional states. While this data can be incredibly valuable in providing customization, personalized insights, and improving our health and well-being, it also poses significant risks if not handled responsibly \citep{felzmann_ethical_2020, kapeller_taxonomy_2020}.

\subsection{Privacy and Data Security}

One of the most pressing concerns surrounding wearable technology is the issue of privacy and data security. As these devices collect and transmit sensitive personal information, it is crucial that robust measures are in place to protect user data from unauthorized access, misuse, or breaches \citep{mone_health_2023}.
At a minimum, wearable technology companies must implement strong encryption protocols, secure authentication methods, and regular security audits to safeguard user data. However, true data privacy goes beyond mere technical solutions. It also requires giving users meaningful insights into how their data are being used as well as control over their own data, with clear and accessible options to view, manage, and delete the information collected by their devices \citep{li_local_2022}.
For example, the European Union's General Data Protection Regulation (GDPR) sets a high standard for data privacy, requiring companies to obtain explicit user consent for data collection and providing users with the right to access and erase their personal data \citep{noauthor_what_2018}. Similar regulations, such as the California Consumer Privacy Act (CCPA), are emerging in other jurisdictions, underscoring the growing global concern for data privacy \citep{noauthor_california_2024}.
However, the challenge remains to make these privacy controls intuitive and user-friendly, so that individuals of all technical backgrounds can easily understand and exercise their data rights. Wearable technology companies must prioritize transparency and clarity in their privacy policies and settings, empowering users to make informed decisions about their personal information \citep{lamb_users_2016}.

\subsection{Algorithmic Bias and Fairness}

As artificial intelligence (AI) becomes increasingly integrated into wearable technology, particularly in health and wellness applications, the issue of algorithmic bias and fairness comes to the forefront. AI systems, when \textit{tuned} or trained on biased or unrepresentative data, can perpetuate or even amplify existing social inequities, leading to disparate outcomes for different user groups \citep{colvonen_limiting_2020}.
For instance, a study by \citet{obermeyer_dissecting_2019} found that a widely used health risk prediction algorithm systematically underestimated the health needs of black patients, leading to inequitable allocation of care resources. An even more direct example can be seen in the tuning of blood-oxygen meters - which fail to identify dangerously low oxygen levels three times more often for individuals with darker skin \citep{fawzy_racial_2022}. Similar disparities are prevalent in a wide range of digital determinants of health (DDoH) - across both technical and medical aspects \citep{charpignon_going_2023}. Such biases, if left unchecked, could lead to wearable technologies that are less effective, unfair, inequitable, or even harmful for certain populations.
To mitigate these risks, it is essential that the datasets used to train AI models for wearables are diverse, inclusive, and representative of the intended user population. This requires active efforts to collect data from underrepresented groups and to test algorithms for fairness across different demographics \citep{caleb_colon-rodriguez_shedding_2023}. Another approach to helping mitigate the design of unjust devices and systems can be found in the Persona Multiplication Method \citep{pearl_persona_2023} where designers systematically examine and identify problems early in the design process.
Moreover, wearable technology companies should adopt ethical AI practices, such as regularly auditing their algorithms for bias, engaging with diverse stakeholders throughout the development process, and being transparent about the limitations and potential biases of their AI systems \citep{maria_trovato_wearables_2023}. By proactively addressing issues of algorithmic fairness, we can ensure that the benefits of AI-driven wearables are equitably distributed.

\subsection{Legal and Regulatory Challenges}

The rapid advancement of wearable technology has often outpaced the development of legal and regulatory frameworks to govern their use. This has created a complex landscape where issues of data protection, health and safety, and consumer rights intersect \citep{bouderhem_privacy_2023}.
Navigating data protection regulations in wearable technology presents a complex challenge, as legal requirements vary across jurisdictions and sectors. In addition to overarching frameworks like GDPR (General Data Protection Regulation) in Europe and CCPA (California Consumer Privacy Act) in the U.S., companies must comply with sector-specific laws such as HIPAA (Health Insurance Portability and Accountability Act), which imposes strict safeguards for personal health information \citep{bouderhem_privacy_2023}.
The complexity deepens when wearables are designed for children or other legally protected groups, where additional privacy safeguards—such as COPPA (Children’s Online Privacy Protection Act) in the U.S.—introduce stricter data collection, consent, and security requirements. Failure to navigate these nuanced regulations can lead to legal repercussions, ethical concerns, and erosion of user trust, making compliance not just a legal necessity, but a foundational aspect of responsible wearable innovation.
Compliance with these regulations is not only a legal necessity but also a matter of maintaining user trust. Data breaches or mishandling of personal information can severely damage a company's reputation and erode user confidence in the technology \citep{cybersecurity_insiders_how_2024}.
Beyond data protection, wearable technologies that collect health data or provide medical advice may be subject to additional regulatory oversight. For example, the U.S. Food and Drug Administration (FDA) regulates certain wearables as medical devices, requiring them to meet stringent safety and efficacy standards \citep{health_overview_2024}.
Navigating this complex regulatory landscape requires not only legal compliance but also a strong ethical compass. Wearable technology companies should establish robust ethical guidelines that go beyond mere legal requirements, addressing issues such as user autonomy, informed consent, and the social implications of their products \citep{tu_ethical_2021}.
By prioritizing ethical considerations alongside legal compliance, wearable technology developers can foster trust, mitigate risks, and ensure that their innovations truly serve the public good \citep{kapeller_implementing_2021}. This requires ongoing dialogue between technologists, policymakers, ethicists, and the public to develop adaptive, responsive frameworks that keep pace with the rapid evolution of wearable technology.
As wearable devices become increasingly enmeshed in our lives, collecting ever more intimate data about our bodies and behaviors, it is imperative that we confront the ethical, social, and legal challenges they present. By prioritizing user privacy, algorithmic fairness, user-control, transparency, and responsible innovation, we can harness the immense potential of wearable technology to improve our health, well-being, and quality of life, while safeguarding the fundamental rights and dignities of individuals in an increasingly connected world.

\section{Future Trends and Innovation}

As wearable technology continues to evolve at a rapid pace, it is crucial to explore the emerging trends and innovations that are shaping the future of this dynamic field. From the integration of cutting-edge technologies like AI, IoT, and AR/VR to the growing emphasis on personalization and sustainability, the landscape of wearables is undergoing a profound transformation. In this section, we will explore these exciting developments, examining how they are redefining the capabilities, user experiences, and environmental impact of wearable devices.

\subsection{Integration with Emerging Technologies}

One of the most significant trends in wearable technology is the seamless integration with a host of emerging technologies, including artificial intelligence (AI), the Internet of Things (IoT), augmented and virtual reality (AR/VR), and blockchain. These integrations are transforming wearables from standalone devices into interconnected nodes within a broader digital ecosystem, unlocking new possibilities for enhanced functionality and smarter interactions \citep{dwivedi_blockchain_2024,xie_integration_2021}.
Smartwatches, for instance, are increasingly designed to interact with smart home systems, allowing users to control various aspects of their living environment directly from their wrist. Imagine being able to adjust your home's lighting, temperature, and security settings with a simple tap on your smartwatch - or simply by arriving home. This seamless integration creates a more cohesive and convenient user experience, where wearables serve as central hubs for managing everyday activities \citep{wicaksono_application_2023}.
In the realm of augmented reality, AR glasses are emerging as powerful tools for providing contextual information and enhancing our perception of the world around us. Picture yourself wearing a pair of AR glasses that display real-time navigation cues, overlaying directions onto the physical world as you walk or drive. Or imagine having instant translations of foreign language text appear right before your eyes, breaking down language barriers and facilitating global communication. These capabilities not only offer unprecedented convenience but also have the potential to greatly enhance safety and accessibility in various settings \citep{chang_a2fitnessartificial_2023}.
Blockchain technology is also making its way into the wearable space, particularly in applications related to data security and user privacy. By leveraging the decentralized and immutable nature of blockchain, wearable devices can offer secure methods for storing and sharing sensitive personal data, giving users greater control over their information and reducing the risk of data breaches \citep{dwivedi_blockchain_2024,xie_integration_2021}. This approach aligns with the growing demand for transparency and trust in digital interactions, ensuring that wearables remain secure and user-centric.

\subsection{Personalization and Customization}

Another major trend driving the future of wearable technology is the increasing emphasis on personalization and customization. As users seek devices that cater to their unique needs, preferences, and lifestyles, wearables are evolving to offer tailored experiences that go beyond one-size-fits-all solutions \citep{kantharaju_framework_2023}.
Generative AI is playing a pivotal role in enabling deep customization in wearables. By leveraging the power of machine learning algorithms, wearable devices can adapt to individual user behavior, preferences, and even physiological characteristics \citep{fu_generative_2025}. Imagine a fitness tracker that not only monitors your physical activity but also learns your specific daily habits and exercise routines and then provides personalized coaching based on your goals, and suggests optimal rest and recovery periods tailored to your body's unique needs.
Customization extends beyond software and user interfaces to encompass the physical design of wearables as well. Modular designs allow users to swap out components, such as watch faces, bands, or even sensors, to create a device that aligns with their aesthetic preferences and functional requirements \citep{botezatu_design_2022}. 3D printing technology is also opening up new avenues for bespoke wearables, enabling the creation of custom-fit devices that are tailored to an individual's unique body shape and size \citep{zhan_3d_2025}.
By embracing personalization and customization, wearable technology is moving beyond mere utility to become an extension of the user's identity and lifestyle. This trend is expected to continue as users demand devices that are not only functional but also deeply personal and reflective of their individuality.

\subsection{Sustainability and Lifecycle Management}

The rise of wearable technology has brought undeniable convenience, but it has also introduced a growing environmental problem. Every new smartwatch, fitness tracker, or smart ring that ends up in a landfill contributes to electronic waste, energy consumption, and carbon emissions. And with rapid product cycles—new models launching every year—most wearables are built to be replaced, not repaired. The industry is increasingly recognizing the need for sustainable practices, prompting a transition toward eco-conscious materials, modular designs, and extended product life cycles \citep{gurova_sustainable_2020}.
A more responsible approach to wearables starts at the design level. Using recycled plastics, biodegradable polymers, and organic textiles can reduce the raw material footprint \citep{liu_eco-friendly_2021}. But materials are only part of the equation. Modular design is arguably an even bigger game-changer—allowing users to repair, upgrade, and swap out components rather than throwing away an entire device when one part fails \citep{breimann_exploiting_2023}. Think of a smartwatch with a replaceable battery, upgradable sensors, or a customizable casing. Instead of being obsolete in two years, it could adapt and evolve, significantly extending its useful life \citep{schischke_impact_2019}.
Beyond better design, companies are experimenting with new business models that prioritize reuse over waste. Some brands are shifting toward product-as-a-service, where users lease or subscribe to a wearable instead of buying it outright. In this model, manufacturers remain responsible for maintenance, refurbishment, and recycling—giving them a strong incentive to build durable, long-lasting devices rather than disposable gadgets \citep{trevisan_circular_2021}.
The pressure is not just coming from inside the industry—consumers are demanding change. As awareness of sustainability grows, many are looking for wearables that align with their values. They want devices that are repairable, recyclable, and ethically produced—not ones that contribute to mountains of e-waste. The companies that recognize this shift and design for longevity rather than obsolescence will be the ones leading the next wave of innovation.
Wearables are no longer just about convenience and connectivity; they are becoming a statement about how we consume technology. The challenge ahead is clear: Can we build wearables that are not only smart and stylish but also sustainable and built to last? The answer will determine whether wearables continue to enhance our lives—or simply become another layer of disposable tech cluttering the planet.

\section{Conclusion}
Wearable technology is no longer just about connectivity or convenience—it has become a deeply integrated part of daily life, shaping how we monitor our health, interact with our environments, and even define personal identity. As we have explored throughout this paper, designing wearables that are truly impactful requires far more than technical innovation alone. Success depends on a multidimensional approach, one that blends engineering, user-centered design, ethical responsibility, and sustainability-driven innovation \citep{ferraro_designing_2011}.
From the earliest fitness trackers to today’s AI-powered, sensor-rich, adaptive wearables, the field has demonstrated a remarkable capacity for transformation. The seamless integration of wearables into daily life has created new opportunities for health monitoring, behavioral insights, and augmented human capabilities. Yet, with this rapid evolution comes an even greater responsibility. As these technologies become more personal—collecting, analyzing, and responding to real-time biometric and behavioral data—the ethical, social, and environmental implications cannot be ignored \citep{bouderhem_privacy_2023}.
Technological advancement alone is not enough. No matter how sophisticated the hardware or how intelligent the algorithms, wearables will fail if they do not align with human needs, values, and behaviors. As we have seen, the most successful wearable technologies emerge not in isolation but from deep, interdisciplinary collaboration. Engineers, designers, behavioral scientists, ethicists, and policy makers must work together to ensure that wearables are not just functional, but intuitive, adaptive, and designed with user trust at their core \citep{fu_generative_2025}.
Privacy, fairness, and transparency must guide the future of wearables, particularly as AI and data-driven personalization become more embedded in these devices. With wearable sensors now capable of detecting everything from heart arrhythmias to emotional states, ensuring secure data practices, clear consent mechanisms, algorithmic accountability, and implementing transparent AI governance is paramount. Without this, wearables risk becoming intrusive rather than empowering \citep{trevisan_circular_2021}.
Sustainability also demands attention. The industry’s prevailing model—where wearables are replaced rather than repaired—exacerbates e-waste and resource consumption. A shift toward modular design, repairability, and circular business models is essential, not just as an environmental obligation but as an opportunity to create longer-lasting, more valuable, and more ethically responsible products that resonate with the needs and aspirations of their users \citep{schischke_impact_2019}.
The future of wearable technology is not predetermined—it is being shaped in real time by the choices of designers, researchers, and users. The next generation of wearables will be defined not just by what they can do, but by how well they integrate into human lives, how responsibly they are developed, and how thoughtfully they address ethical and environmental concerns. Those who push beyond mere technological progress—who center their work on adaptability, privacy, sustainability, and human experience—will define the future of this industry.
Wearables have the potential to empower, enhance, and transform the way we live. But the true measure of their success will not be in their processing power or feature sets—it will be in whether they make life better, more meaningful, and more connected, without compromising trust, agency, or sustainability. As we continue this journey, the challenge is clear: to ensure that wearable technology is not just innovative, but responsible; not just advanced, but human-first.
The future of wearables is being built now. The question is, what kind of future will we choose to create?

\bibliographystyle{unsrtnat}
\bibliography{main}  

\end{document}